%% file: manuscript.tex
\documentclass[prl,nofootinbib,floatfix,twocolumn,showpacs,superscriptaddress]{revtex4}
\usepackage[dvips]{pstcol}
\usepackage{graphicx,amsmath}
\usepackage{dcolumn}
\usepackage{bm}
\usepackage{rotating}
\usepackage{amsmath,amssymb,amsfonts}
\usepackage{color}	

\newcommand{\emcal}{EMCal}

\definecolor{grey}{rgb}{0.75,0.75,0.75}
\definecolor{orange}{rgb}{1.0,0.5,0.5}
\definecolor{brown}{rgb}{0.5,0.25,0.0}
\definecolor{pink}{rgb}{1.0,0.5,0.5}
\definecolor{green}{rgb}{0.,0.5,0.}

\begin{document}



\title{
Performance of prototypes for the ALICE electromagnetic calorimeter}

\input{authors}

\date{\today}

\begin{abstract}
The performance of prototypes for the ALICE electromagnetic sampling calorimeter has 
been  studied in test beam measurements at FNAL and CERN.
A $4\times4$ array of final design modules showed an energy resolution 
of about 11\% /$\sqrt{E(\mathrm{GeV})}$ $\oplus$ 1.7 \% with a uniformity of the response
to electrons of 1\% and a good linearity 
in the energy range from 10 to 100 GeV. 
The electromagnetic shower position resolution was found to be described by
1.5 mm $\oplus$ 5.3 mm  /$\sqrt{E \mathrm{(GeV)}}$.
For an electron identification efficiency of 90\% a hadron rejection factor of $>600$ 
was obtained.
\end{abstract}


\maketitle

%
\section{Introduction}
%
ALICE (A Large Ion Collider Experiment) at the LHC is designed to carry 
out comprehensive measurements of
high energy nucleus--nucleus collisions, in order to study QCD matter
under extreme conditions and to explore  the phase transition between confined
matter and the Quark-Gluon Plasma (QGP)~\cite{CH1Ref:PPRvol1, CH1Ref:PPRvol2}. 

ALICE contains a wide array of detector systems for measuring hadrons, leptons
and photons. 
The ALICE detector is described in detail in~\cite{ALICEjinst}.
The large acceptance Electromagnetic Calorimeter (EMCal), 
which is foreseen to be fully installed in 2011,
significantly enhances ALICE's capabilities for jet measurements. 
The ALICE \emcal\ is designed to provide the following functions:

-- efficient and unbiased fast level L0 and L1 trigger on high energy jets 

-- measurement of the neutral portion of jet energy

-- improvement of jet energy resolution

-- measurement of high momentum photons, $\pi^0$ and electrons

-- $\gamma$ / $\pi^0$ discrimination up to 30 GeV~\footnote{Considering 
invariant mass and shower shape techniques only.}

-- e / h separation (for momenta larger than 10 GeV/c)

-- high uniformity of response for isolated electromagnetic clusters.

From Monte Carlo simulations,
a detector energy resolution of the order of about $15\%/\sqrt{E \mathrm{(GeV)}} \oplus 2\%$ 
was found to be
sufficient for the jet physics program and is fixed as the minimum detector requirement.
The electron and photon physics programs, however, would benefit from better resolution. 

The overall design of the EMCal is heavily influenced by its integration
within the ALICE~\cite{ALICEjinst} setup
which constrains the detector acceptance to a region of about 110 degrees in azimuth
$\phi$, $-$0.7 $<$ $\eta$ $<$ 0.7 in pseudo-rapidity and 
4.35 m $< \mathrm{R}_{\mathrm{EMCal}} < 4.7$ m radial distance .
\\\newline\indent
This paper presents the performance of prototype modules
studied in test beam measurements at FNAL and at CERN.
The goals of these measurements were the determination of the intrinsic energy and position
resolution, the investigation of the linearity and uniformity of the detector response,
the determination of the light yield per unit of energy deposit and a study of
the response to electrons and hadrons.
Furthermore, monitoring and calibration tools were successfully implemented and tested.
%
\section{Calorimeter module design and readout}
%
The chosen technology is a layered Lead(Pb)-Scintillator(Scint) sampling calorimeter 
with wavelength shifting (WLS) fibers that run longitudinally through the Pb/Scint
stack providing light collection (Shashlik)~\cite{EMCal-TDR:2008}. 
The basic building block is a module consisting of $2\times2$ optically isolated
towers which are read out individually; each spans $\Delta \eta \times
\Delta \phi = 0.014 \times 0.014$ each.
White, acid free, bond paper serves as a diffuse reflector on the scintillator surfaces and
provides friction between layers. 
The scintillator edges are treated with TiO$_2$ loaded reflector to improve the transverse
optical uniformity within a single tower and to provide tower to tower
optical isolation better than $99\%$.

The energy resolution for a sampling 
electromagnetic calorimeter varies with the sampling frequency approximately
as $\sqrt{d_{\rm Sc}/f_{\rm s}}$, where $d_{\rm Sc}$ is the
scintillator thickness in mm and $f_{\rm s}$ is the sampling fraction for
minimum ionizing particles (MIPs). For optimum resolution in a given physical space
and total radiation length, there is thus a desire to have the highest
possible sampling frequency.
Practical considerations, including the cost
of the total assembly labour, suggest reducing the total number
of Pb/Scint layers thus decreasing the sampling frequency.
The requirement of a compact detector consistent with the EMCal integration volume
and the chosen detector thickness of about 20 radiation lengths, results in a
lead to scintillator ratio by volume of about 1:1.22 corresponding to a sampling geometry of
Pb(1.44 mm)/Scint(1.76 mm).

The physical characteristics of the EMCal modules are summarized in Table~\ref{Table-1}.
\begin{table}[t]
\begin{center}
\caption{EMCal module physical parameters. Here, RL stands for Radiation Length and MR for the 
Moliere Radius.} 
\begin{tabular}{ll}
\hline
Parameter & Value \\ \hline 
Tower Size (at $\eta$=0) & $\sim$6.0 $\times$ $\sim$6.0 $\times$ 24.6 cm$^3$ \\  
Tower Size & $\Delta \phi \times \Delta \eta = 0.0143 \times 0.0143$ \\ 
Sampling Ratio & 1.44 mm Pb / 1.76 mm Scint. \\ 
Layers & 77 \\  
Scintillator & Polystyrene (BASF143E $+$ \\
             & 1.5\%pTP $+$ 0.04\%POPOP)  \\ 
Absorber  & natural Lead \\
Effective RL X$_0$ & 12.3 mm \\ 
Effective MR R$_M$ & 3.20 cm \\  
Effective Density & 5.68 g/cm$^3$ \\  
Sampling Fraction & 1/10.5           \\  
Radiation Length & 20.1 \\  \hline
\end{tabular}
\label{Table-1}
\end{center}
\end{table}
%
%
%

Scintillation photons produced in each tower are captured by an array of 36
Kuraray Y-11 (200 M), double clad, wavelength shifting (WLS) fibers.
Each fiber within a given tower terminates in
an aluminized mirror at the front face of the module and is integrated
into a polished, circular group of 36 fibers at the photo sensor end at the back of
the module.
The 6.8 mm diameter fiber bundle from a given tower connects to the 
Avalanche Photodiode (APD)
through a short light guide/diffuser.
The selected photo sensor is the Hamamatsu S8664-55 avalanche photodiode
chosen for operation in the high field inside the ALICE magnet.
The APDs are operated at moderate gain for low noise and high gain stability in order 
to maximize energy and timing resolution. 
The number of primary electrons generated in the APD by an electromagnetic
shower is $\approx 4.4$ electrons/MeV.   
The reverse bias voltage of the APDs are individually controlled to provide
an electron multiplication factor (M) of 30 resulting in a charge output of  
$\approx 132$ electrons/MeV from the APDs.
All APDs used for the test beam measurements were previously calibrated~\cite{APD-test}.
The charge output from the APD is integrated by a Charge Sensitive Preamplifier (CSP)
with a short rise time of $\approx 10$ ns and a long decay time of $\approx 130~\mu s$, 
i.e., approximately a step pulse.
The amplitude of the step pulse is proportional to the number of integrated electrons 
from the APD and therefore  proportional to the energy of the incident particle.
The output from the CSP is conditioned with a second order gaussian shaper in order to make
the signal suitable for digitization with the Alice TPC Readout Chip~\cite{altro-chip}.

The readout electronics of the PHOS (PHoton Spectrometer) detector~\cite{CH3Ref:PHOS}
of ALICE have been adopted for the EMCal front end electronics readout with
only minor modifications, as the light yield per unit of
energy deposit in the EMCal is
similar to that of the PHOS~\cite{CH3Ref:PHOS2}.
A detailed description of the EMCal (PHOS) front end electronics (FEE) and their
performance is given in Ref.~\cite{CH3Ref:PHOSFEE}. 
The FEE has an effective 14-bit dynamic range over the interval 16 MeV to 250 GeV
resulting in a Least Significant Bit on the low gain range of 250 MeV (10-bits) 
and on the high gain range of 16 MeV. 
Compared to PHOS, the coarse granularity of the \emcal\ yields higher occupancies.
As the number of read out samples recorded is dictated by the total shaped pulse width,
a shaping time of 200 ns (2 $\mu$s for PHOS) is chosen in order to keep the total 
data volume per central unit similar to PHOS and to fulfill the constraints from 
the total available bandwidth. 
This results in an electronic noise contribution of about 12\;MeV per EMCal tower.
However, due to the larger intrinsic energy resolution term of EMCal compared to 
PHOS the importance of the electronics noise contribution is much less.
The effect of the shaping time on the calorimeter resolution has been studied 
in the test beam measurements performed at FNAL and is discussed in the next 
section.
%
\section{Test beam Measurements}
\label{sec:test-beam}
%
The performance of the first ALICE EMCal modules constructed according to final 
design was studied in CERN SPS and PS test beam lines in autumn 2007.
The test utilized a stacked $4\times4$ array of
EMCal modules ($8\times8$ towers).
All towers were instrumented with the full electronics chain with shapers and
APD gains operated as planned in ALICE.
A LED calibration system was installed in order to
monitor time-dependent gain changes.
The readout of the front end electronics used the standard ALICE 
data aquisition system.

Earlier test measurements were performed in November 2005 at the Meson Test Beam (MTEST)
at FNAL
utilizing a stacked $4\times4$ array of prototype EMCal modules ($8\times8$ towers)
of slightly different design than the final one,
such as a radiation length of 22 $X_0$ and a sampling geometry of
Pb(1.6~mm)/Scint(1.6~mm).
For this test in particular, measurements were made for
comparison of the performance with two different signal shaping times in the front end electronics.
Two front end electronics cards (32 towers each) were used for the readout of the modules; 
one had the
nominal $2~\mu$s signal shaping time of the PHOS, and the
other had a modified 200~ns shaping time as planned for EMCal.

The goals of the test beam measurements were:
\begin{itemize}
   \item To determine the intrinsic energy and the position resolution using electron beams.
  \item To investigate the linearity and uniformity of the response; in particular across towers
and module boundaries and for tilted or recessed modules.
\item To determine the light yield (signal) per unit of deposited electromagnetic energy.
\item To study the effect of shorter shaping times as planned for the final design.
\item To study the energy dependence of the response to electrons and
hadrons to determine the particle identification
capabilities of the EMCal.
\item To develop and investigate the performance of the monitoring and calibration tools (gain stability, 
time dependencies)
using electron beams, MIPs from hadron beams, LED events and cosmic muons.
\item To develop and test ALICE standard software for readout, calibration and analysis.
\end{itemize}
%
\subsection{Test setup and beam line instrumentation}
%
The characteristics of the test beams at FNAL and CERN are summarised in Table
~\ref{tab:test-beams}.
\begin{table}[ht!]
\begin{center}
\caption{Test beam parameters. }
\begin{tabular}{llll}
\hline
Lab & FNAL & CERN \quad\quad & CERN \\ \hline 
test beam & MT6 & SPS H6 & PS T10 \\
particle & e,h & e,h & e,h \\
intensity [s$^{-1}$]\qquad\quad & $10^3$-$10^4$ & $10^2$-$10^3$ & $10^2$-$10^3$ \\
$\Delta p/p$ & $\pm$ 1\% &  $\pm$ 1.3\% & - \\ 
P$_\mathrm{range}$ [GeV] & 3-33 &  5-100 & 0.5-6.5 \\ 
purity & mixed beam \qquad & $>99 \%$ & mixed beam \\ \hline 
\end{tabular}
\label{tab:test-beams} 
\end{center}
\end{table}

For handling and stacking purposes, the modules were assembled on a
strong-back in strip units of four modules in the vertical direction.
In order to scan the entire surface of all four modules they were placed on a
remotely controlled movable platform.
The range of both horizontal and vertical adjustment allowed to scan the
whole array of modules. 

\begin{figure}[t]
\centering
\rotatebox {0}{\includegraphics *[width=1.05\columnwidth] {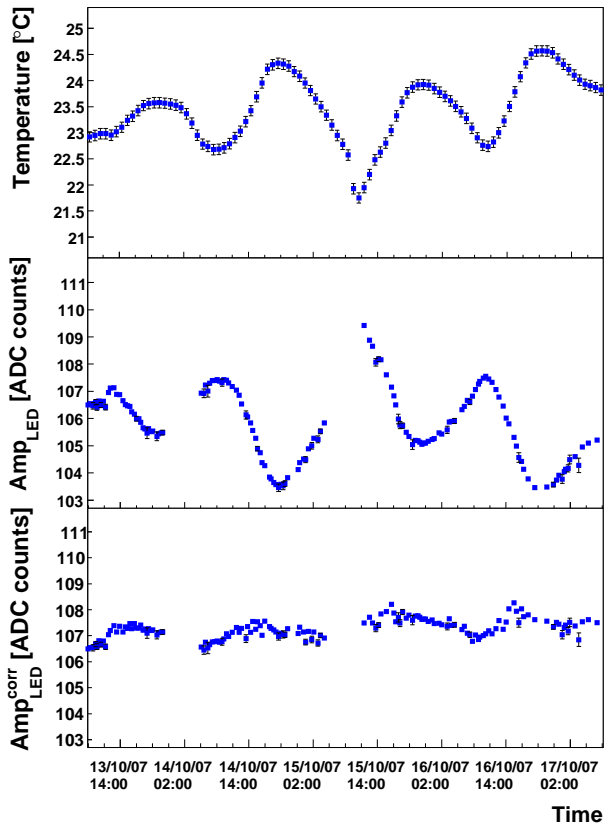}}  
\caption{Upper panel: temperature measurement as a function of time.
Mid panel: LED signal amplitude for a typical tower for the same time interval.
Lower panel: corrected LED signal amplitude.
}
\label{figure:LEDvsTemp}
\end{figure}
The EMCal readout electronics were
attached to the back of the array of modules with the
electronics cards and readout units located on the same
moveable table as the modules, together with the low voltage supplies.
In both setups at CERN and FNAL,
a pair of scintillator paddles upstream of the EMCal was used for the beam definition trigger.
In addition, at the CERN PS and at the FNAL MTEST, the signals from gas threshold 
$\hat{\mathrm{C}}$erenkov counters
were used as an electron trigger for electron/pion discrimination.
A set of three MWPCs in front of the EMCal provided $x-y$ position measurements with better than
 1~mm position resolution for the setup at FNAL. 
The MWPCs were used to define the beam particle trajectory which could then be
projected to the front face of the EMCal modules.

The official ALICE data aquisition (DATE v6.13)~\cite{CHARef:date}
was used for taking the EMCal data.
The MWPC  data was recorded with a CamacCrate-via-USB (CCUSB) readout system.
The data from the $\hat{\mathrm{C}}$erenkov counters
were also recorded via the CCUSB system.
The EMCal data were combined with the data from the trigger
detectors and from the
MWPCs offline, aligning the information from the different data streams spill-by-spill.
\begin{figure}[t]
\centering
\rotatebox {0}{\includegraphics *[width=1.05\columnwidth] {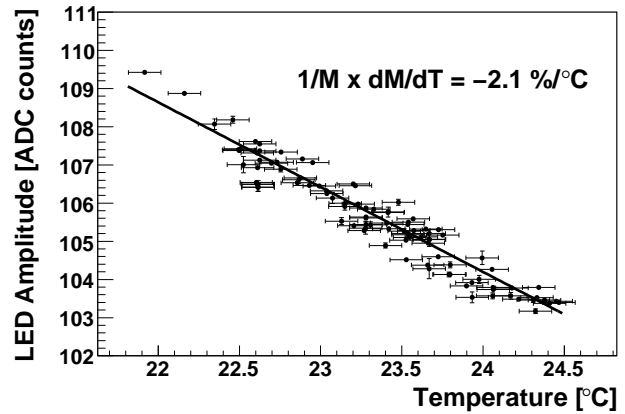}}
\caption{LED signal amplitude as a function of the measured temperature.}
\label{figure:TempCoeff}
\end{figure}
%
\subsection{LED calibration system}
\label{sec:calib}
%
In order to reach the design EMCal energy resolution for high energy electromagnetic 
showers, a tower-by-tower relative energy calibration of better 
than 1\% has to be obtained and maintained in the offline analysis.
In addition, since analog tower energy sums provide the basis of the level L0 and L1
high energy shower trigger input to the ALICE trigger decision, the EMCal should
operate with APD gains adjusted to match online relative tower energy calibrations
to better than 5$\%$.

A LED calibration system, in which all towers view a calibrated pulsed LED
light source, has been successfully tested to track and adjust for the temperature
dependence of the APD gains during operation.
The LED triggers were collected in parallel with the beam particle events throughout the
entire CERN test beam measurements. These measurements were performed with
the APDs operated at the nominal fixed M=30 gain.

The variation of the EMCal response to the LED signal with time and temperature was studied in order
to test the system for calibration purposes.
The temperature was monitored by a total of  eight temperature sensors installed on 
the back surface of the module.
The measured LED signal amplitude variation for a given tower as a function of time is compared in
 Fig.~\ref{figure:LEDvsTemp}  for the same time interval with the temperature readings
from the nearest sensor for the module in which the tower was located.
A clear anti--correlation is observed.

Over the whole data taking period, some sharp variations in the  LED signal amplitude were observed
that cannot
be attributed to temperature changes but rather to LED light yield changes, as when the setup
was reconfigured.
These changes of the overall LED light were taken into account with an iterative extraction of
the temperature coefficients.
First, a new time interval was defined if an APD amplitude changed by
more than 20\,\% from one hour to the next. For each time interval, both low and high gain
LED signal amplitudes were fit simultaneously as planes in space
defined by signal amplitude, temperature, and the time interval.
In a first iteration, all points deviating by more than $1.5\,\sigma$ from a predefined
slope range ($0.015<\left|\mathrm{d}M/\mathrm{d}T\right| [\%/^\circ\mathrm{C}]<0.025$) were excluded.
In the next iteration, the cleaned sample was fit with a free parameter for the slope
in order to define the temperature coefficient.
 Fig.~\ref{figure:TempCoeff}
shows the LED amplitude for a typical tower as a function of the temperature and for
a certain time interval.
The temperature coefficients obtained from the fits of these distributions 
were used to correct for the time dependence of the APD gain.
As an example, the corrected LED amplitude is shown in the lower panel of  Fig.~\ref{figure:LEDvsTemp}
for the considered time interval.

The selected
LED event amplitudes as well as the information from the temperature sensors as a function
of time are stored in a database.
An interface was developed and tested that allows for time-dependent calibration corrections in the
offline analysis of the test beam data.
%
\subsection{Signal reconstruction}
%
The digitized time samples from the read out have an amplitude as a function of
time $t$ that can be described with the form of a $\Gamma$-function in $ADC(t)$, where
\begin{eqnarray}
\label{eq:Sig}
ADC(t) = {\rm pedestal + A}\cdot e^{-n} \cdot x^\mathrm{n} \cdot  e^{\mathrm{n} \cdot (1-x)}\; ,   \\
x = (t - t_0) / \tau\; .        \nonumber
\end{eqnarray}
Here, $\tau=\mathrm{n}\cdot\tau_0$ with the shaper constant $\tau_0$
 and n=2
as the shaper is gaussian of second order
(composed of a differentiator and two integrators~\cite{CH3Ref:PHOSFEE}).
The charge collected
from the APD, and hence the energy deposited in the tower, is proportional
to  the value of the \mbox{parameter $\rm A$} at  the time value 
$(t_0+\tau)$ where the function peaks.

The test beam data were used to investigate the performance of this function 
and the parameters were optimized.
The High-Low gain correlation was studied using the electron data
in order to determine a threshold value for the amplitude for which the low gain
rather than the high gain needs to be used due to saturation (at 1023 ADC counts).
A good  High-Low gain correlation with an average ratio of 16.3 between both gains was found
up to 1050 ADC counts 
when using the values from a fit for ADC counts $>$ 1000.

\begin{figure}[t]
\centering
\rotatebox {0}{\includegraphics *[width=0.95\columnwidth]{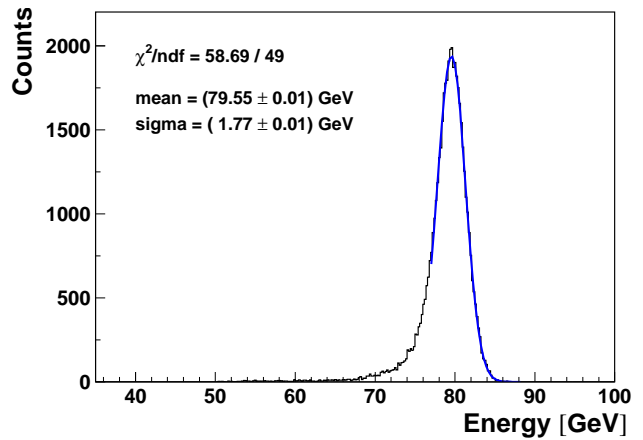}}
\caption{
Reconstructed energy for 80~GeV incident electrons. The curve represents 
a fit of a truncated gaussian to the histogram with fit results as given
in the figure.  
}
 \label{figure:energyReco}
\end{figure}

An overall inter-calibration procedure was carried out for all
towers by normalizing the hadron MIP amplitudes in each tower,
to a reference tower. Isolation of the MIP peak was
achieved requiring, for each tower, no energy deposit in the
surrounding eight towers.
An alternative inter-calibration map was
also considered by using the information given by the
electron beam peak in each tower.
An absolute calibration for each tower was accomplished by comparing the nominal electron beam
 energy with the corresponding peak in the energy spectrum, as obtained by a
sum over a $3\times3$ tower cluster. For this purpose, $3\times3$ local
cluster inter-calibration coefficients were extracted from the
overall map, by choosing each tower in turn as a reference. This
allowed evaluation of the energy spectrum by a sum over the 9 towers
in the cluster, with a proper calibration adjusted to match that of
the central tower in each cluster.
These calibration coefficients were used to analyse the test beam data with 
the standard ALICE cluster reconstruction software.
Fig.~\ref{figure:energyReco} shows the reconstructed energy for 80\;GeV
incident electrons (for a typical run).
%
%
\subsection{Linearity and uniformity of energy response}
%
%
\begin{figure}[t]
\rotatebox {0}{\includegraphics *[width=0.90\columnwidth]{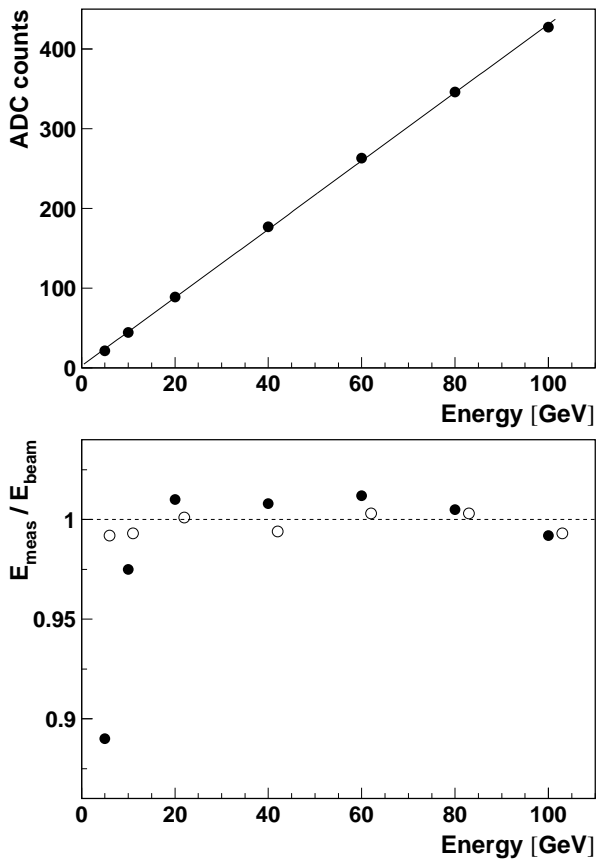}}
\caption{
Top: Linearity of the response for a sum over a 3$\times$3 tower cluster as a function
of the incident beam energy. The solid line is a linear fit to the data.
Bottom:
Ratios between the linear fit and the data (full circles) and a cubic fit and
the data (open circles).
The dashed line is placed at a ratio of unity to guide the eye.
}
 \label{figure:linearity}
\end{figure}
The absolute energy calibration obtained by a sum over a 3$\times$3 tower cluster
is shown in Fig.~\ref{figure:linearity}.
The linearity of the response is better than 1\% over the full energy range
down to 20\;GeV.
At low energies, threshold effects become non--negligible compared to the total energy deposited
and light transmission losses might have an impact.
In fact, as shown in the bottom panel of Fig.~\ref{figure:linearity} by the full circles
the reconstructed energy is systematically lower than the incident one for energies
equal or below 10\;GeV. A drop of $\sim10\%$ is observed at 5\;GeV.
This behaviour is well described by a cubic function as 
demonstrated by the open circles in Fig.~\ref{figure:linearity} (bottom panel).
At high energies, deviations of the ratio from unity are expected due to longitudinal 
shower leakage.
The data show an indication of such an energy loss at high energies.

The uniformity of the energy response was studied for several different conditions.
All module centers and a major part of tower centers were scanned using 80\;GeV electrons.
In addition, data were taken across tower and module borders.
A uniformity of the energy response was found with a RMS better than 1\;GeV,
for 80\;GeV incoming electrons.
This result implies a very good uniformity of the response (within 1\%) 
for the EMCal as constructed.
%
\subsection{Light Yield}
%
The light yield, the number of photoelectrons at the APD per
unit of electromagnetic energy deposited in the EMCal
(photoelectrons/MeV), determines the overall APD+shaper gain 
required to match the desired dynamic range in ALICE.
Due to the large number of individual towers planned for the final
design of the EMCal, it is also important to estimate the
tower-to-tower dispersion of the light yield.

During the test beam, the APDs were all operating at gain M=30.
The individual voltage settings had been established 
for each APD prior to the test beam measurements.
This procedure compares the amplitude at a given bias
voltage to the amplitude measured at low voltages, where the gain is
assumed to be unity~\cite{APD-test}.
 The light yield (LY) at the gain M=30, for each
individual tower, is then extracted following
\begin{eqnarray}
\label{eq:LY}
LY(p.e./MeV) &=& (channels/MeV) \cdot (1/G_A)  \\
 & & \cdot (1/P_G) \cdot (1/ADC_{conv}), \nonumber
\end{eqnarray}
where the shaper amplifier gain $G_A$ = 0.229, the charge voltage
conversion factor of the preamplifier $P_G$ = 0.83 V/pC and the ADC
conversion $ADC_{conv}$ = 1024 channels/V.
The light yield at unit gain (M=1) is obtained from this value
divided by 30. 
An average light yield of about (4.3 $\pm$ 0.3) photoelectrons/MeV,
was found, which is consistent with the light yield value of PHOS.
%
\subsection{Energy resolution}
\label{sec:test-beam-resolution}
%
The energy resolution of an electromagnetic calorimeter can be parameterized as
\begin{center}
\begin{eqnarray}
\sigma(E) / E = a \oplus b / \sqrt{E}  \oplus c / E\; ,
\label{eqn:resolution-fit}
\end{eqnarray}
\end{center}
where E is the measured energy.
The intrinsic resolution is characterized by the parameter
$b$ that arises from stochastic fluctuations due to intrinsic detector effects such
as energy deposit, energy sampling, light collection efficiency, etc. 
The
constant term, $a$, originates from systematic effects, such as shower leakage,
detector non-uniformity or channel-by-channel calibration errors. 
The third
term, $c$, is due to electronic noise summed over the towers of the cluster
used to reconstruct the electromagnetic shower.
The three resolution contributions add together in quadrature.

Detailed GEANT3 Monte Carlo simulations for the final module design yield
fit results using Eqn.(\ref{eqn:resolution-fit}) of $a=(1.65 \pm 0.04)\%$, 
$b = (8.0 \pm 0.2)\%$ and $c = (7.4 \pm 0.2)\%$ over a photon energy range of 0.5 GeV  to 200 GeV.
These results are
based on energy deposition only and do not include photon
transport efficiencies.
Systematic
contributions to the resolution arising from calibration and related
systematic uncertainties are ignored.
The value of the constant term $a$ is
dominated by longitudinal shower leakage in these calculations.
Other systematic effects,
which arise during detector fabrication and from the tower-by-tower
calibration uncertainties, will increase $a$.
%
%

\begin{figure}[t]
\centering
\rotatebox {0}{\includegraphics *[width=0.95\columnwidth]{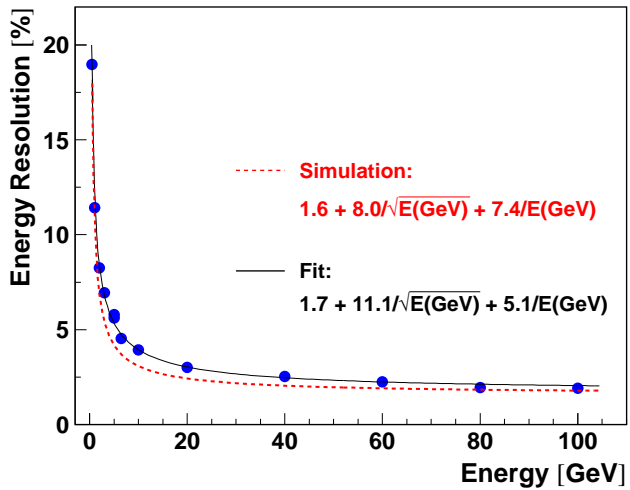}}
\caption{Energy resolution for electrons as a function of the incident beam momentum.
The beam energy spread was subtracted from the measured result.
The dashed curve represents the resolution obtained from Monte Carlo simulations. 
}
\label{figure:resVsMomentum-CERN}
\end{figure}

By combining data taken at the CERN PS and SPS the calorimeter resolution over the
energy range of 0.5\;GeV to 100\;GeV could be explored.
Such energy scans were performed at several different positions, including tower and
module edges.
The LED calibration system was used to track and adjust for the time dependence of
the calibration coefficients.
No systematic variation of the resolution depending on the position was observed.
The resolution obtained at the different positions was combined and the
average values as a function of the incident beam momentum are displayed
in Fig.~\ref{figure:resVsMomentum-CERN}.
The momentum spread of the incident beam of typically 1.3$\%$ was subtracted in
quadrature.
A fit to the energy resolution as a function of the
incident energy following Eqn.(\ref{eqn:resolution-fit}) 
is also shown in Fig.~\ref{figure:resVsMomentum-CERN} with
parameters $(a = 1.7 \pm 0.3)\%$, $(b = 11.1 \pm 0.4)\%$ and $(c = 5.1 \pm 0.3)\%$.
These parameters can be compared with the GEANT3 simulation result
for the EMCal module geometry described before and presented by the dashed line 
in Fig.~\ref{figure:resVsMomentum-CERN}.
The increase of the stochastic term $b$, representing a worse intrinsic resolution 
compared to the Monte Carlo simulations, is mainly due to light attenuation and 
light collection inefficiencies which were not modelled. 
The small increase of the constant term $a$ demonstrates
a stable, high quality detector fabrication and a good tower-by-tower
calibration. 
The linear term, modelling electronic noise contributions, is set too 
high in the simulation.

\begin{figure}[t]
\centering
\rotatebox {0}{\includegraphics *[width=0.95\columnwidth] {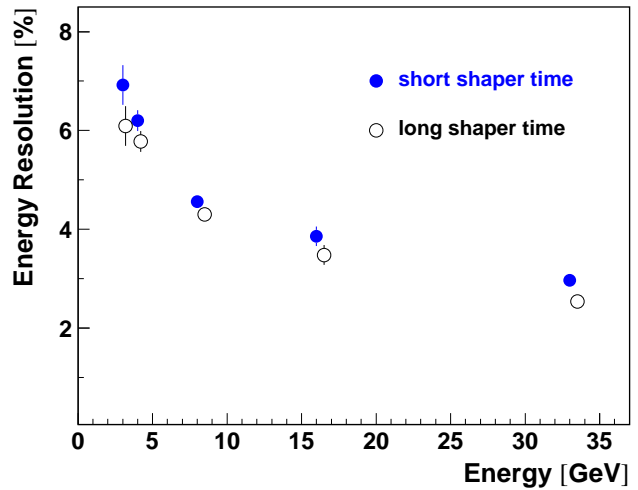}}
\caption{Energy resolution for electrons as a function of the incident beam momentum for short 
(full circles) and long (open circles) shaper time, corresponding to the 
EMCal and PHOS design, respectively. The open circles are slightly shifted to the right 
for visibility.
}
\label{figure:resVsMomentum-FNAL}
\end{figure}
The energy resolution was also studied for different incidence locations corresponding
to the modules as installed in ALICE.
Most of the test beam data were taken with
a configuration where the beam hits the EMCal modules perpendicularly, corresponding
to $z=0, \eta$=0 position.
Data were also taken with configurations where the modules were tilted in $\phi$ 
by 6 or 9 degrees
at different surface positions.
The energy resolution for such tilted configurations 
compares well with the average resolution as a function of energy presented
in Fig.~\ref{figure:resVsMomentum-CERN}.
No significant deviations from the average resolution at zero degree was observed.
\\
\newline
Using the data from the FNAL test beam, possible effects of the shorter design shaping 
time for the EMCal of 200~ns (compared to 2~$\mu$s for PHOS) were studied.
Fig.~\ref{figure:resVsMomentum-FNAL} shows the energy resolution as a function of the
incident energy.
The results are shown separately for the short (full circles) and long (open circles) shaping time readout
regions of the test setup, averaged over various runs in each region. 
The resolution slightly deteriorates when using the short shaping time
but is still well within the detector requirements.
%
%
\subsection{Position resolution}
%
%
\begin{figure}[t]
\centering
\rotatebox {0}{\includegraphics *[width=0.95\columnwidth]{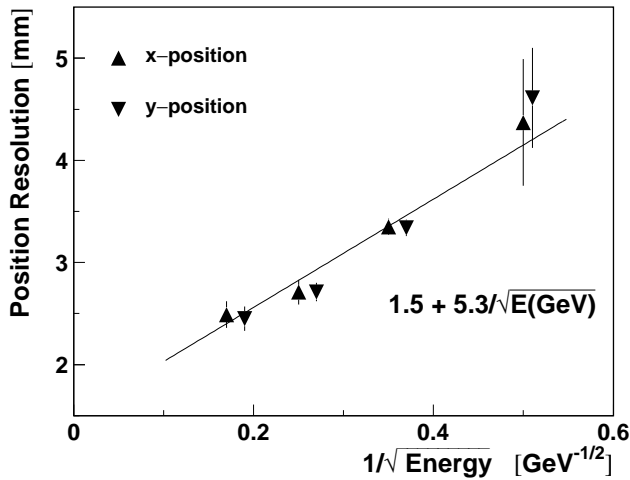}}
\caption{Dependence of the position resolution as a function of the deposited energy for
electrons. The curve shows the best fit result.
The triangles representing the resolution for the y-position are slightly shifted 
to the right for visibility.
}
\label{figure:PosResEDep}
\end{figure}
%
%
The segmentation of the calorimeter allows one to obtain the hit position from
the energy distribution inside a cluster with an accuracy better than the 
tower size.
The $x$ and $y$ coordinate locations are calculated using a
logarithmic weighting~\cite{CHARef:LOG} of the tower energy deposits. 
Data from the FNAL test beam were used where the MWPCs provided a
reference position measurement of better than 1mm.
Fig.~\ref{figure:PosResEDep} shows the $x$ and $y$ position resolution
as a function of the energy deposit for electrons.
As expected, no difference in the resolution in the $x$ and $y$ directions is observed. 
The electromagnetic shower position resolution is seen to be described as 
1.5 mm $\oplus$ 5.3 mm/$\sqrt{E \mathrm{(GeV)}}$, where the two 
contributions add together in quadrature.
%
\subsection{Response to hadrons}
\begin{figure}[t]
\rotatebox {0}{\includegraphics *[width=0.95\columnwidth]{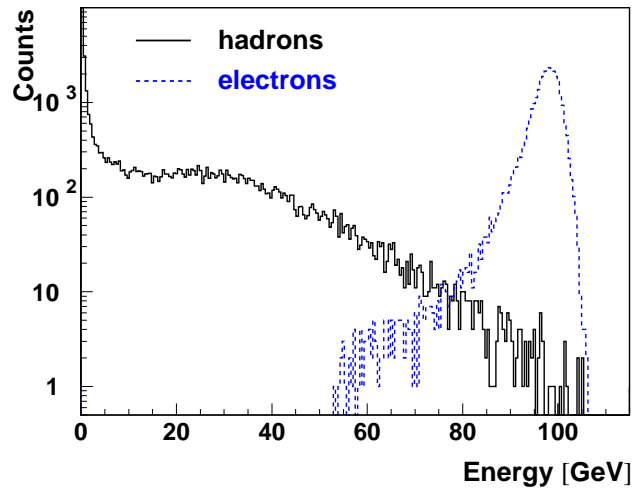}}
\caption{EMCal response to hadrons (full histogram) and electrons (dashed histogram) of 100\;GeV.
}
\label{fig:hadron-electron-response}
\end{figure}

%
The EMCal can further enhance the ALICE particle identification capabilities due to
the characteristically different response to electrons and hadrons.
While electrons leave all their energy in the calorimeter, hadrons 
leave only a fraction of their energy but show a long tail due to hadronic
showers.
At the CERN PS pure electron and hadron beams were available.
Fig.~\ref{fig:hadron-electron-response} shows the reconstructed
energy for an electron and hadron beam of 100~GeV,
which illustrates this very distinct response to
electrons and hadrons.
The high energy tail in the hadron response originates from processes such as
charge exchange $\pi^- + p \rightarrow \pi^0 + n$, where most of 
the energy of the charged pions goes into neutral pions. 
These neutral pions decay immediately into photons starting a cascade which is 
indistinguishable from an electron--initiated shower. 

\begin{figure}[t]
\rotatebox {0}{\includegraphics *[width=0.95\columnwidth]{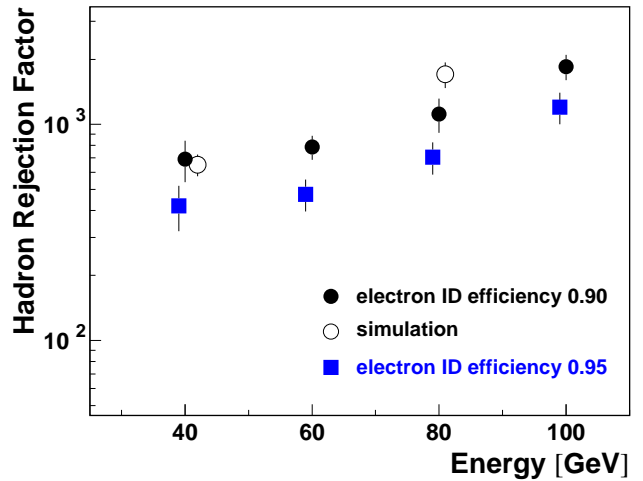}}
\caption{
Hadron rejection as a function of the incident hadron beam energy for an electron
identification efficiency of 90\% (circles) and 95\% (squares).
Error bars represent the total uncertainty. 
The open circles show the result from a Monte Carlo simulation for 90\% 
electron identification efficiency.
The squares (open circles) are shifted to the left (right) for visibility.
}
\label{fig:HRF}
\end{figure}

The hadron rejection factor is defined as the number of all hadrons divided by the number
of hadrons misidentified as electrons. 
This factor is shown in Fig.~\ref{fig:HRF} as a function of
the incident hadron beam energy for electron identification efficiencies of 90\% and 95\%.
Error bars give the total uncertainty, which is dominated by the systematic
uncertainty of the evaluation.
Results from a Monte Carlo simulation are also shown for an
electron identification efficiencies of 90\%.
For each incident beam energy,
the electron identification efficiency was determined by integrating the reconstructed energy 
distribution of the pure electron beam  (dashed histogram in Fig.~\ref{fig:hadron-electron-response})
from the right-hand side till a cut value corresponding to the chosen efficiency.

A rejection factor of $10^2$ to $10^3$ is obtained over the energy range of 40~GeV to 100~GeV.
Test beam data at lower hadron energies were not taken.
Hadron/electron rejection can be further improved by considering the characteristic 
shower shapes, as hadrons produce showers with wider spatial distributions than electrons.
%
%
\subsection{Cosmic ray calibration}
%
A calibration of all modules will be performed before their insertion in ALICE. 
The calibration procedure is based on a measurement of cosmic-ray muons
at the minimum of ionization.

The muon signal measured in each tower is obtained by the use of an isolation 
procedure applied offline.
For each event, the maximum signal amplitude is chosen and for all neighboring towers
a signal  smaller than a threshold value required.
This threshold value is limited by the electronic noise 
(set to 3 ADC channels in the
present case which amounts to about 15\% of the muon energy).

Since the energy of MIP muons is too low to trigger the EMCal, an external trigger 
is necessary. 
The muons, passing the towers along their length, are selected using scintillator paddles.
Each paddle covers 
12 modules grouped into a `strip module', and is read out at both extremities by
photomultiplier tubes (PMT). 
This trigger configuration appeared to be the most reliable from the cosmic 
analysis test done in December 2007 ~\cite{EMCal-TDR:2008} with the EMCal prototype described 
above.
The time of flight difference between both PMTs allows one to  select vertical
muons with a spatial accuracy of a few centimeters. 
The isolation procedure then ensures that no energy was deposited in
the neighboring towers. 
A 24-hour run allows the accumulation of about 500 muons 
per towers, which is sufficient to extract a
MIP peak with an accuracy better than 1\%.

\begin{figure}[t]
\hspace{-0.3 cm}
\rotatebox {0}{\includegraphics *[width=0.95\columnwidth]{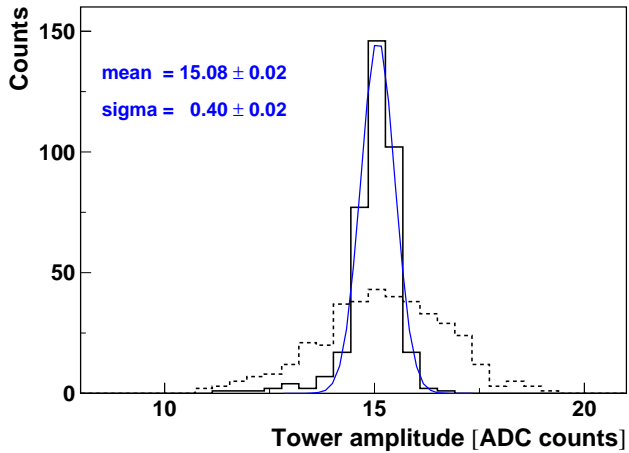}}
\caption{
Response of 384  towers of the EMCal to cosmic-ray muons before (dashed histogram)
and after (full histogram) individual gain calibration.
The curve represents a fit of a gaussian to the full histogram with fit results as given
in the figure.  
}
\label{fig:cosmics}
\end{figure}

An individual gain calibration is performed for each tower, so as to ensure that 
the amplitude of the average signal for cosmic
muons is the same for all towers. 
The tower gains, which are controlled through the tower high voltage power, are tuned
iteratively. 
Fig.~\ref{fig:cosmics} shows the dispersion of the mean amplitude of 384 towers 
before and after this procedure (thin and bold lines, respectively). 
After three iterations a final relative dispersion $<$ 3\% is reached.
%
%
\section{Conclusion}
%
The performance of a ($4\times4$) array of prototype modules and of a ($4\times4$) array of
final design modules for the ALICE \emcal\ has been studied in test beam 
measurements at FNAL and CERN, respectively.
\\
These studies demonstrate:
(i) an average light yield of (4.3 $\pm$ 0.3) photoelectrons/MeV

(ii) an energy resolution of 
\begin{eqnarray}
\frac{\sigma(E)}{E} [\%]  = (1.7\pm0.3) \oplus \frac{(11.1\pm0.4)}{\sqrt{E(\mathrm{GeV})}} \oplus \frac{(5.1\pm0.7)}{E(\mathrm{GeV})}  \nonumber
\label{CH4:eq1}
\end{eqnarray}

(iii) a uniformity of the response within 1\% for all towers and configurations

(iv) a good linearity of the response to electrons in the energy range 10-100 GeV

(v) an only slightly deteriorated energy resolution when using the EMCal default shaping
time of 200 ns compared to 2 $\mu$s for PHOS 

(vi) a position resolution described by 1.5 mm $\oplus$ 5.3 mm/$\sqrt{E{\mathrm{(GeV)}}}$

(vii) a hadron rejection factor $>$ 600 for an electron identification efficiency of 90\%.
\\\newline 
A LED calibration system was successfully tested to track and adjust for 
temperature dependent effects during operation.

Cosmic ray calibrations allow a precalibration of all modules prior to installation in ALICE
with a relative spread of $<$ 3\%, sufficent for the use in an online trigger.

\indent The performance of the tested \emcal\ modules reaches all design criteria.
%
\section{Acknowledgments}
%
We gratefully acknowledge the CERN and the FNAL accelerator devisions for
the good working conditions in the testbeam facilities.
We would like to thank all engineers and technicians of the participating 
laboratories for their invaluable contribution to the construction of the
EMCal.
This work was supported by
the Helsinki Institute of Physics and the Academy of Finnland;
the French CNRS/IN2P3/INPG, the 'Region Pays de Loire', 'Region Alsace','Region Auvergne' and CEA, France;
the Istituto Nazionale di Fisica Nucleare (INFN) of Italy; 
the Office of Nuclear Physics within the United States DOE Office of Science
and the United States National Science Foundation.

\end{document}

%% file: authors.tex

\def\groupfrascati{\affiliation{Laboratori Nazionali di Frascati, INFN, 00044 Frascati, Italy}}
\def\groupCataniaInfn{\affiliation{Sezione INFN, 95123 Catania, Italy}}
\def\groupCataniaUni{\affiliation{Dipartimento di Fisica e Astronomia dell'Universit\`a Catania e Sezione INFN, 95123 Catania, Italy}}
\def\groupRoma{\affiliation{Museo Storico della Fisica e Centro Studi e Ricerche Enrico Fermi, 00184 Roma, Italy}}
\def\groupCern{\affiliation{CERN, European Organization for Nuclear Research, 1211 Geneva, Switzerland}}
\def\groupLPSC{\affiliation{LPSC, Universit\'e Joseph Fourier Grenoble 1, CNRS/IN2P3/INPG, Institut Polytechnique de Grenoble,  38026 Grenoble Cedex, France}}
\def\groupSubatech{\affiliation{SUBATECH, Ecole des Mines de Nantes, Universit\'e de Nantes, CNRS-IN2P3, 44307 Nantes, Cedex 3, France}}
\def\groupIPHC{\affiliation{IPHC, Institute Pluridisciplinaire Hubert Curien, Universit\'e Louis Pasteur, CNRS-IN2P3, 67037 Strasbourg, Cedex 2, France}}
\def\groupJyvaskyla{\affiliation{Helsinki Institute of Physics (HIP) and University of Jyv\"askyl\"a, 40351 Jyv\"askyl\"a, Finland}}
\def\groupValenciaA{\affiliation{Valencia Polytechnic University, 46022 Valencia, Spain}}
\def\groupValenciaB{\affiliation{University of Valencia, 46010 Valencia, Spain}}

\def\groupBerkeley{\affiliation{Lawrence Berkeley National Laboratory, Berkeley 94720, United States}}
\def\groupCalifornia{\affiliation{California Polytechnic State University, San Luis Obispo 93407, United States }}
\def\groupCreighton{\affiliation{Creighton University, Omaha Nebraska 68178, United States }}
\def\groupHouston{\affiliation{University of Houston, Houston 77204, United States}}
\def\groupTennessee{\affiliation{University of Tennessee, Knoxville 37996, United States}}
\def\groupLivermore{\affiliation{Lawrence Livermore National Laboratory, Livermore 94550, United States}}
\def\groupOakridge{\affiliation{Oak Ridge National Laboratory, Oak Ridge 37831, United States}}
\def\groupWayne{\affiliation{Wayne State University, Detroit 48202, United States}}
\def\groupYale{\affiliation{Yale University, New Haven 06511, United States}}

\def\groupCentralChina{\affiliation{Hua--Zhong Normal University, 430079 Wuhan, China}}



\groupBerkeley
\groupCataniaUni
\groupCataniaInfn
\groupWayne
\groupfrascati
\groupCern
\groupLPSC
\groupHouston
\groupJyvaskyla
\groupTennessee
\groupLivermore
\groupSubatech
\groupYale
\groupOakridge
\groupCreighton
\groupRoma
\groupCalifornia
\groupIPHC
\groupValenciaB
\groupValenciaA
\groupCentralChina


\author{J.~Allen}  \groupfrascati
\author{T.~Awes}  \groupOakridge
\author{A.~Badal\'a}  \groupCataniaInfn
\author{S.~Baumgart\footnote{now at: RIKEN, The Institute of Physical and Chemical Research, Wako 351-0198, Japan}}  \groupYale
\author{R.~Bellwied}  \groupWayne
\author{L.~Benhabib}  \groupSubatech
\author{C.~Bernard}  \groupLPSC
\author{N.~Bianchi}  \groupfrascati
\author{F.~Blanco}  \groupCataniaUni \groupHouston
\author{Y.~Bortoli}  \groupSubatech
% ...
